\documentclass[journal=nalefd, manuscript=letter]{achemso}
\usepackage[version=3]{mhchem} 
\usepackage[T1]{fontenc}       

\author{Ewa Rej}
\author{Torsten Gaebel}
\author{David E.J. Waddington}
\author{David J. Reilly}
\affiliation{ARC Centre of Excellence for Engineered Quantum Systems, School of Physics, University of Sydney, Sydney, NSW 2006, Australia.}
\email{david.reilly@sydney.edu.au}

\title
  {Hyperpolarized Nanodiamond Surfaces}

\abbreviations{HP,NMR,ESR, ND, AO, DNP, MR, MRI, OE, SE, HPHT}

\begin{document}







\begin{abstract}
The widespread use of nanodiamond as a biomedical platform for drug-delivery, imaging, and sub-cellular tracking applications stems from their non-toxicity and unique quantum mechanical properties. Here, we extend this functionality to the domain of magnetic resonance, by demonstrating that the intrinsic electron spins on the nanodiamond surface can be used to hyperpolarize adsorbed liquid compounds at room temperature. By combining relaxation measurements with hyperpolarization, spins on the surface of the nanodiamond can be distinguished from those in the bulk liquid. These results are likely of use in signaling the controlled release of pharmaceutical payloads. 
\end{abstract}

\section{Introduction}

Bio-functionalized nanoparticles are emerging as highly versatile platforms upon which to develop the new theranostic and tailored imaging modalities needed in the era of personalized medicine \cite{NPinmed, Mura_Couvreur}. These nanoscale agents, smaller than most subcellular structures, open the prospect of detecting and imaging a spectrum of diseases with enhanced sensitivity, and offer a means of targeting the delivery and controlled release of pharmaceutical payloads \cite{Wang_Thanou,  Li}. Enabling such advanced applications will require a detailed understanding of the chemical interface of a nanoparticle  in vivo, configuring its complex interaction with, for instance, the extracellular matrix, disease processes, or the tumor microenvironment. 

Magnetic resonance (MR) techniques are well-suited for probing bio-chemical reactions involving nanoparticles in vivo, but challenging in the limit of small concentrations, where interactions at the nanoparticle interfacial surface lead to only fractional changes in the dominant signal arising from the surrounding fluid \cite{toolbox1, snsno}. The difficulty in isolating signals that are derived specifically from the nanoparticle surface has led to new techniques based on hyperpolarization to enhance the sensitivity of MR spectroscopy, mostly via the use of surface-bound radicals \cite{DNPsens, mollevelchar, Rossini1, DNPsens3, DNPsens2, piveteau2015structure}.  These techniques have been extended to dynamic nuclear polarization (DNP) of liquids using both extrinsic \cite{hypso, songi} and intrinsic \cite{Cassidy_PRB} nanoparticle defects. In the case of nanodiamond (ND), a biocompatible nanoscale allotrope of carbon \cite{huang, Liu_nanotechnology}, the rich surface chemistry \cite{kruger} is ideally suited to binding small molecules such as proteins, ligands, antibodies, or therapeutics, making it a promising substrate for loading and targeted delivery applications \cite{zhang, Chow,huang, chen, adsorption, giammarco2016adsorption}. Further, cellular imaging and tracking of NDs is also possible using fluorescent color centers \cite{Say_review, mcguinness}, and future developments may combine imaging with nanoscale magnetic and electric fields sensing capabilities \cite{maze, balasubramanian}. Complementing these atributes, the detection of hyperpolarized $^{13}$C nuclei in the ND core \cite{us, casabianca, dutta} has recently opened the prospect of new MRI modalities based on nanodiamond. 

Here, we demonstrate that $^1$H nuclear spins from liquid-state compounds including water, oil, acetic acid, and glycerol mixtures can be hyperpolarized using X-band microwaves at room temperature via contact with free-electron impurities on the surface of nanodiamond. Rather than the Overhauser mechanism usually seen in the polarization of liquids, we observe a DNP frequency spectrum that is indicative of the solid-effect, in which the polarized $^1$H spins are those that are adsorbed on the nanodiamond surface. Further, by combining low field ($B$ $<$ 1 T) spin relaxation measurements and hyperpolarization we demonstrate that it is possible to determine the extent to which the ND surface is saturated by its liquid environment. In combination with modalities based on hyperpolarized $^{13}$C in the ND core \cite{us}, these results are likely of use in enabling in vivo approaches to monitor the binding and release of biochemicals from the functionalized ND surface.

\section{Results}

\subsection{Nanodiamond Surfaces}
The nanodiamonds used in these experiments are manufactured using the high pressure high temperature (HPHT) technique and  purchased from Microdiamant. A micrograph showing NDs with an average size of 125 nm is shown in Fig. 1a. Measurements were made on diamonds in a size range between 18 nm and 2 $\mu$m. Adsorption of the compounds onto the ND surface occurs passively when diamonds are mixed and sonicated with various liquids. 

\begin{figure}
\includegraphics{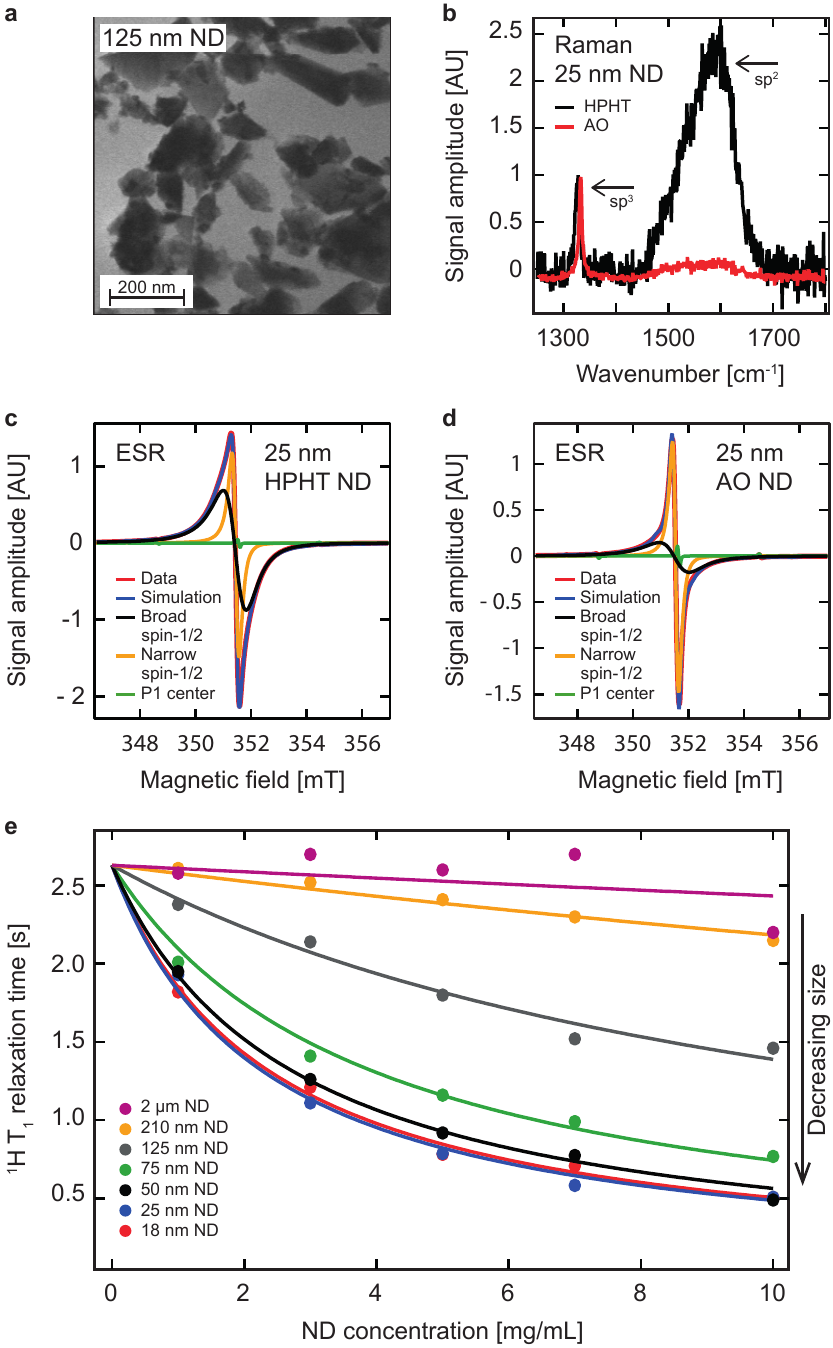}
\caption{\small \textbf{Characterizing ND surfaces.} {\bf a)} Electron micrograph of 125 nm ND. {\bf b)} Comparison of Raman spectra for HPHT ND (black) and air oxidized (AO) ND (red). Raman spectra show sp$^2$ hybridized carbon from the surface of the diamond and sp$^3$ hybridized carbon from the core of the diamond. The sp$^2$ Raman cross section is $\sim$150 times larger than the sp$^3$ Raman cross section leading to a comparatively larger peak. The fluorescence of the diamond has been subtracted using a baseline correction, and spectra have been normalized to the sp$^3$ hybridized peak. {\bf c,d)} Comparison of the ESR spectrum of 25 nm HPHT ND and 25 nm AO ND. The data (red) is simulated (blue) using three components: a narrow spin-1/2 Lorentzian component (yellow), a broad spin-1/2 Lorentzian component (black) and a P1 center component (green). {\bf e)} The $^1$H T$_1$ relaxation time of water in water-ND mixtures as a function of ND size and concentration at $B$~=~330~mT. Data points are fits to the $^1$H T$_1$ build up performed using an inversion recovery sequence. The solid lines are fits to the relaxivity equation [see methods section]. Smaller NDs (25 nm ND, blue dots, relaxivity: $R$\,=\,0.17\,mg$^{-1}$\,mL\,s$^{-1}$) have a larger effect upon the T$_1$ relaxation time of water than larger NDs (2~$\mu$m~ND, purple dots, relaxivity: $R$ = 0.003 mg$^{-1}$ mL s$^{-1}$).}
\end{figure}

Using Raman spectroscopy, we observe that our NDs comprise two phases of carbon, sp$^2$ hybridized, attributed to carbon on the surface of the ND, at wavenumber $\nu$ \nobreak = \nobreak 1580 \nobreak cm$^{-1}$, and sp$^3$ hybridized, attributed to carbon in the core of the ND, at $\nu$ \nobreak = \nobreak 1332 \nobreak cm$^{-1}$, as shown in Fig. 1b. The sp$^2$ carbon phase results in free electrons and provides a surface for liquid adsorption. We observe more sp$^2$ hybridized carbon in smaller NDs than larger NDs, due to the much higher surface to volume ratio. Air oxidation \cite{Oxidized} of the NDs etches away some of the surface, removing sp$^2$ hybridized carbon and surface electrons. 

Much of the functionality of ND, including its fluorescence, magnetic field sensitivity, and use as an MRI contrast agent stems from the presence of impurities and unbound electrons in the crystal lattice or nanoparticle surface. For DNP applications these intrinsic free-radicals provide a means of hyperpolarizing nuclear spins \cite{us}, but also open pathways for spin relaxation \cite{Furman}. For the smaller NDs, the dominant electronic defects are carbon dangling bonds on the surface, contributing a  broad spin-1/2 component in an electron spin resonance (ESR) spectrum, (see black trace in Fig. 1c). Air oxidation of the NDs removes some of these surface electrons, resulting in a decrease in the broad spin-1/2 component in the spectrum, as shown in Fig. 1d. Other components of the ESR spectrum include a narrow spin-1/2 component (yellow), attributed to defects in the core of the ND and a P1 center component (green) which results from a substitutional nitrogen atom with the electron hyperfine coupled to the $^{14}$N spin. Increasing the diameter of the NDs shrinks the surface to volume ratio and reduces the relative amplitude of the broad and narrow spin-1/2 components. At the same time,  the larger NDs have more P1 centers in the core.

We first examine whether the presence of these free electron spins on the diamond surface can be identified by mixing the nanoparticles with various liquids containing $^1$H spins at $B \sim$ 330 mT. In this configuration, the presence of free electrons on the ND surface enhances the spin relaxation (with characteristic time $T_1$) of the surrounding  $^1$H  from the liquid, as shown for the case of water in Fig. 1e (and Supplementary Figs. 1-2). Consistent with the ESR measurements, we find that this relaxivity effect is more prominent for small NDs, which have a larger surface to volume ratio, and a higher number of surface spins.  We note that although the relaxivity effect is small when compared to commonly used contrast agents based on metal conjugates \cite{Manus2009Gd}, it is significant enough to enable $T_1$-weighted imaging when using concentrations of order 1mg/mL. 

\subsection{Nanodiamond as a Hyperpolarizing Agent}

Turning now to a key result of the paper, we make use of room temperature hyperpolarization as a means of further probing and identifying the spins at the liquid-nanodiamond interface. In contrast to high-field hyperpolarization modalities that aim to increase the MR signal for enhanced contrast, our focus here is the spectrum of the polarization with frequency, enabling different hyperpolarization methods to be distinguished. The Overhauser effect, for instance, is commonly observed when polarizing liquid compounds comprising molecules that undergo rapid translational and rotational diffusion. This mechanism relies on scalar and dipolar relaxation pathways to build up a nuclear polarization when driving at the electron Larmor frequency, $f = \omega_e$, resulting in positive or negative enhancement depending on the electron-nuclear coupling.  In contrast, if the 
nuclear and electron spins are bound such that the primary mode of nuclear spin relaxation is via the same electrons used for polarizing \cite{Protonchars1, Protonchars2, french}, then hyperpolarization occurs via the solid-effect, cross-effect, or thermal-mixing mechanism, see Fig. 2a and 2b. 

\begin{figure}
\includegraphics{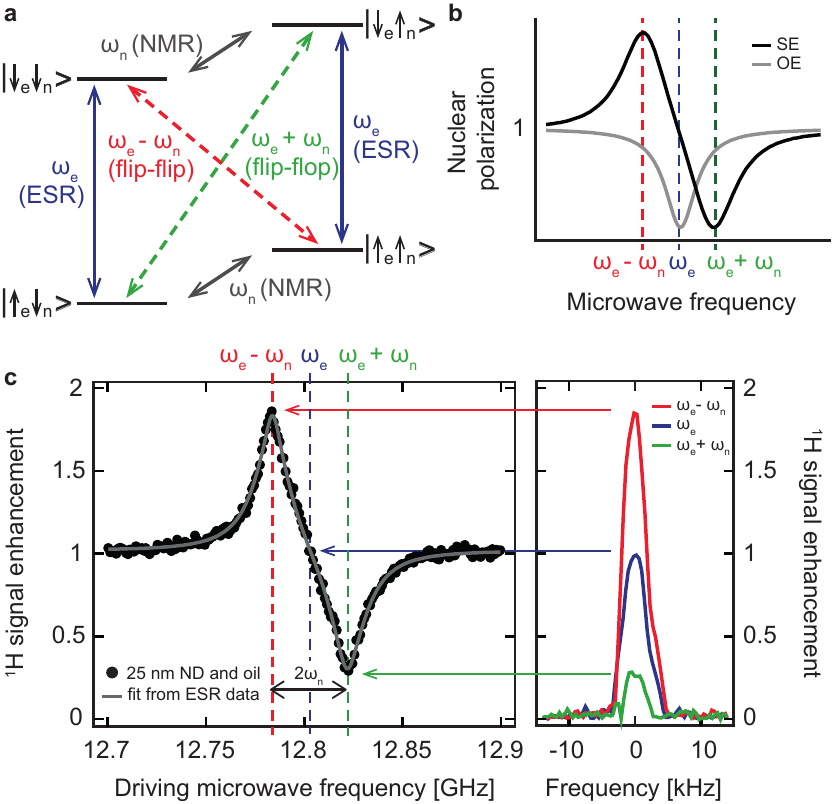}
\caption{\small  \textbf{Solid effect enhancement of adsorbed liquids on NDs.} {\bf a)} Energy level diagram for a dipolar coupled electron and nuclear spin-1/2 system in a magnetic field. The ESR (blue), NMR (grey), flip-flop (green) and flip-flip (red) transitions are shown. For the solid effect, driven flip-flip transitions (red) at a frequency $f = \omega_e - \omega_n$ involve a mutual electron and nuclear flip resulting in a positive nuclear polarization, shown in b.  Driven flip-flop transitions (green) result in a negative nuclear polarization. For Overhauser effect hyperpolarization, saturating the ESR transition can lead to positive or negative enhancement (shown in b), through relaxation via the zero quantum (green) or double quantum (red) transitions respectively. {\bf b)} Theoretical enhancement spectra for the solid effect (black) and Overhauser effect (grey) hyperpolarization mechanisms. {\bf c)} $^1$H signal enhancement as a function of driving microwave frequency at $B$\;=\;458\;mT (black dots). The fit to the data (grey line) is based on the ESR trace linewidths for the broad and narrow spin-1/2 impurities in the ND. The hyperpolarization spectrum is consistent with that given by the solid effect. Enhancement is given by the hyperpolarized signal divided by the signal with microwaves off.}
\end{figure}

Using the naturally occurring electrons on the surface of NDs, we are able to hyperpolarize the $^1$H spins in a range of liquid-nanodiamond compounds including water, oil, acetic acid, and glycerol mixtures, despite their variation in chemical polarity. The data presented in Fig. 2c is representative of the effect, showing in this case $^1$H hyperpolarization from oil (Sigma O1514) mixed with 25 nm ND. The data clearly exhibit the signature of hyperpolarization via the solid-effect, with a positive signal enhancement when driving at $f = \omega_e - \omega_n$ and a negative signal enhancement at $f = \omega_e + \omega_n$. No enhancement is seen at $f$ = $\omega_e$, as would be expected if the Overhauser effect was contributing to the hyperpolarization and there is no enhancement in liquid solutions without NDs. The presence of the solid-effect provides a strong indication that the signal enhancement stems from hydrogen spins that are adsorbed at the nanodiamond surface. Similar behavior is observed when hyperpolarizing other liquid-nanodiamond mixtures  (see Supplementary Fig 3). We also note that the enhancement scales inversely with nanoparticle size (see Supplementary Fig 4.), a further indication that DNP is mediated via spins on the ND surface and consistent with the relaxivity measurements presented in Fig. 1e.

We further examine the hyperpolarization spectrum as a function of magnetic field over the range $B$\,=\,300\;mT\,-\,500\;mT, as shown in Fig 3a. The position of the enhancement peaks follow $f = \omega_e - \omega_n$ and $f = \omega_e + \omega_n$ with a peak splitting of $f$ = 2$\omega_n$ (black dashed lines) at low magnetic fields, (see Fig. 3b). Small deviations from the predicted splitting are evident, potentially due to a contribution from the cross effect or thermal mixing mechanism. A contribution from thermal mixing and the cross effect is expected given the ESR spectrum contains a broad spin-1/2 component that is wider than the nuclear Larmor frequency. Surprisingly, we also observe that the $^1$H signal enhancement increases with magnetic field, as shown in Fig. 3c. This dependence with field is currently not understood, given that the solid- and cross-effect enhancements are expected to scale in proportion to $1/B^2$ and $1/B$ respectively \cite{Maly}. Field-dependent spin relaxation of electrons, as well as a narrowing of the ESR line with increasing field, may lead however, to more efficient hyperpolarization. 

\begin{figure}
\includegraphics{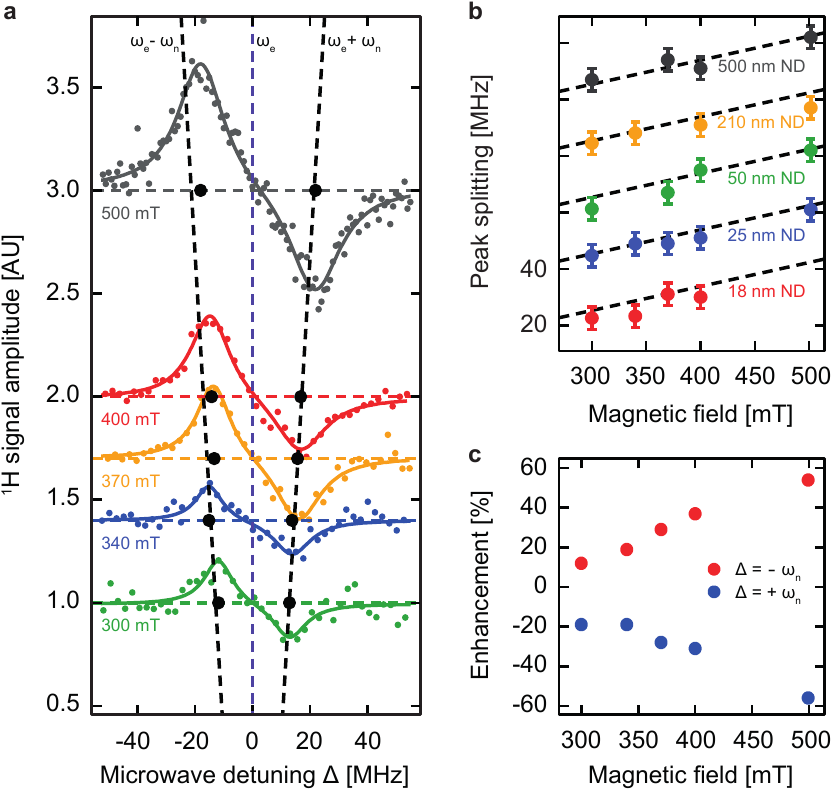}
\caption{\small {\textbf{Hyperpolarization behavior at various magnetic fields.} {\bf a)} The hyperpolarized $^1$H NMR signal in oil adsorbed onto 25 nm ND as a function of driving microwave frequency at magnetic fields between $B$ = 300\;mT and $B$ = 500\;mT ($B$ = 300\;mT in green, 340\;mT in blue, 370\;mT in yellow, 400\;mT in red and 500\;mT in grey). The solid lines are bi-Lorentzian fits to the data [see methods section}]. The positions of the peaks (black dots) follow the lines $f = \omega_e - \omega_n$ and $f = \omega_e + \omega_n$ (black dashed lines). We see no hyperpolarization at $f = \omega_e$ (blue dashed line). The microwave detuning is given by $\Delta$ = $f$ - $\omega_e$. The traces have been offset by the magnetic field scaling for clarity. {\bf b)}\,The frequency splitting between the maximum and minimum   $^1$H signal from oil adsorbed on the ND surface for 18 nm ND (red), 25 nm ND (blue), 50 nm ND (green), 210\,nm\,ND (yellow), and 500 nm ND (grey). The splitting follows the predicted value for the solid effect of $f = 2\omega_n$ (dashed lines). Error bars are 10 \% of the Lorentzian fit to the hyperpolarization data. Traces have been offset for clarity. {\bf c)}  The $^1$H signal enhancement as a percentage of the non-polarized signal at magnetic fields between $B$ = 300\;mT\,-\,500\;mT for a 25 nm ND and oil mixture. Positive enhancement at $f = \omega_e - \omega_n$ is shown in red and negative enhancement at $f = \omega_e + \omega_n$ is shown in blue. Data points are the saturation value of an exponential fit to a polarization build up divided by the signal with far detuned microwaves.}
\end{figure}

\subsection{Hyperpolarization at the Nanodiamond Surface}

Mixing nanodiamond with a significant amount of liquid leads to behavior indicative of a system with two independent spin baths. In our ND-water illustration shown in Fig. 4a, the $^1$H spins in the bulk of the liquid comprise one bath, with the other being those spins that are absorbed on the surface of the nanodiamond, in contact with free electrons. We isolate the independent contribution to the signal from each bath by comparing spin relaxation as a function of water concentration and in the presence of hyperpolarization using the pulse sequences shown in Fig. 4b and 4c. 
Consistent with the behavior expected for two spin baths, the relaxation decay exhibits a bi-exponential dependence with a short $T_1$ and long $T_1$, as shown by the black curve in Fig. 4d. We attribute the short  $T_1$ (green shading) to spins adsorbed on the ND surface, where the presence of electrons can rapidly relax nuclear spins in close proximity. When the ND-water mixture is sufficiently diluted ($>$ 60 $\mu$L),  a longer tail in the decay appears (blue shading) that likely stems from spins in the bulk liquid, decoupled from the ND surface. Reducing the amount of water, Fig. 4e shows that the long time component is suppressed since all spins can then be rapidly relaxed by the ND surface. 

\begin{figure}
\includegraphics{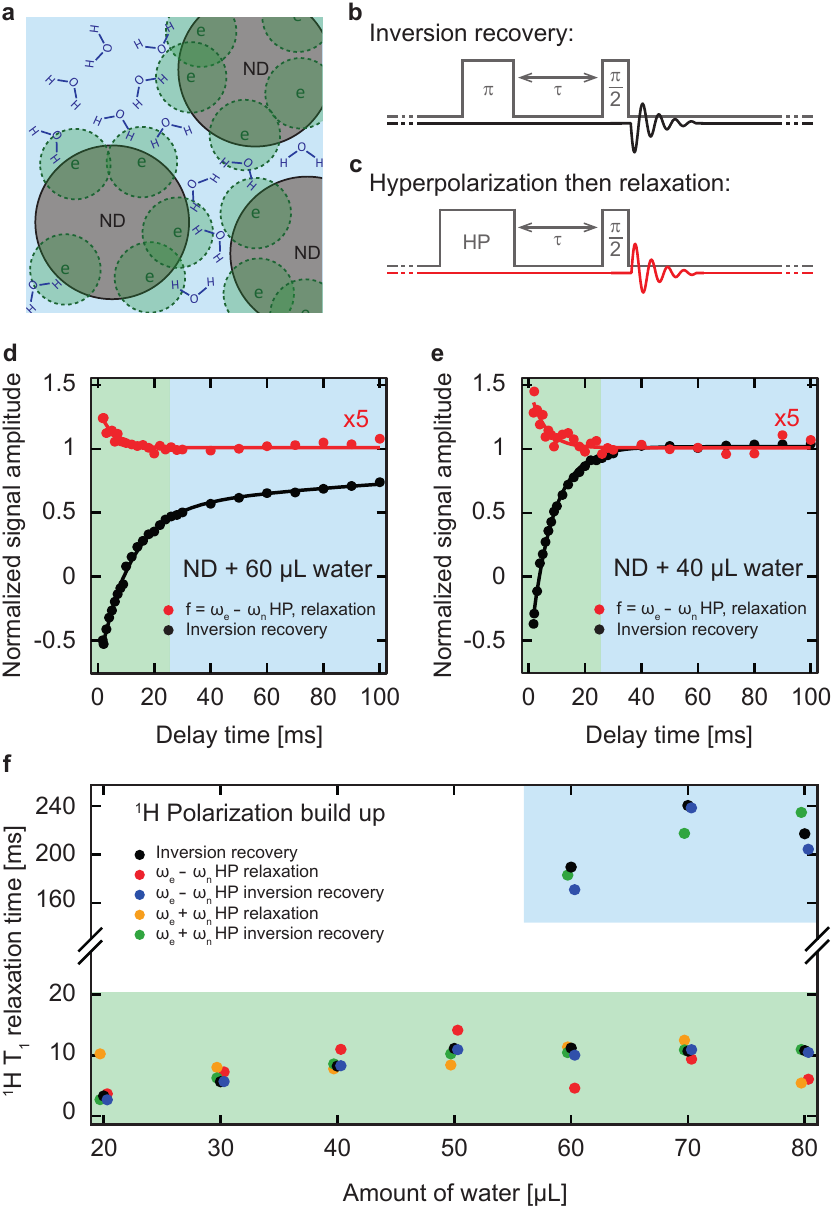}
\caption{\small \textbf{T$_1$ relaxation in hyperpolarized liquids.}  {\bf a)} Schematic of a ND-water mixture. ND is shown in grey and water is shown in blue. Surface electrons (e) on the NDs are shown as a green circle representing the distance over which hyperpolarization occurs. {\bf b, c)} Pulse sequences used to measure relaxation of a hyperpolarized liquid. {\bf b)} Inversion recovery sequences probing both spins in the bulk and adsorbed water. {\bf c)} Hyperpolarization followed by a $\pi$/2 pulse probes the relaxation of spins close to the ND surface. {\bf d, e)} Relaxation of a water-ND mixture measured by the pulse sequences outlined in b, c. {\bf d)} We observe a double exponential behavior in the inversion recovery experiment (black) when sufficient water has been added (60 $\mu$L), indicating two distinct spin baths. The HP relaxation (red) falls off at a fast rate.  {\bf e)} When only a small amount of water is added (40 $\mu$L), we only observe fast relaxation in both the hyperpolarization (red) and inversion recovery (black) experiments. Solid lines are exponential and double exponential fits to the data. The short component of the relaxation is shaded green, and the long component is shaded blue. {\bf f)} Summary of the $^1$H T$_1$ relaxation times in water-ND mixtures as a function of water concentration. We always observe a short component to the relaxation ($\tau_1 \sim 10$\;ms) and we begin to observe a long component ($\tau_2 \sim 200$\;ms) for inversion experiments once $\sim$ 60 $\mu$L of water is added to the ND. Data points are exponential and double exponential fits to five relaxation experiments: inversion recovery (black), positive enhancement then relaxation (red), negative enhancement then relaxation (yellow), positive enhancement then inversion recovery (blue), and negative enhancement then inversion recovery (green).}
\end{figure}

Inserting a hyperpolarization (HP) pulse in place of the usual inversion recovery sequence leads to a small enhancement of the signal that rapidly decays, irrespective of the concentration of water in the system. This behavior, taken together with the relaxation measurements, suggests that hyperpolarization is again limited only to those nuclear spins absorbed on the ND surface, consistent with DNP occurring via the solid-effect. Further, this data puts a bound on the extent to which diffusion can transport hyperpolarized spins from the ND surface to  the bulk of the liquid. Since no signal enhancement is ever observed in the long-time component of the relaxation, we conclude that the hyperpolarized spins that are bound to the surface are unable to diffuse into the bulk before relaxing. In further support of this picture, Fig. 4f shows that relaxation (without hyperpolarization), suddenly switches from a single exponential to bi-exponential decay when the amount of water exceeds 60 $\mu$L (black dots). This behavior is symptomatic of any sequence that contains an inversion recovery pulse (black, green, or blue dots), since the pulse acts on both the spins on the surface and those in the bulk (see Supplementary Figs. 5 and 6). Hyperpolarization without inversion recovery however, acts only on ND-surface spins and always leads to fast, single-exponential decay independent of the amount of water.

A complimentary picture emerges when nanodiamond is mixed with oil. We again conclude that the system comprises two distinct baths with different spin dynamics, in this instance by examining how the signal is enhanced by hyperpolarization beyond the thermal contribution, $\Delta$S = S$_{\textrm{HP}}$ - S$_{\textrm{Th}}$, shown in red in Fig. 5a. In the limit of no oil, Fig. 5b shows that there is no enhancement in the signal since the electrons on the ND surface cannot hyperpolarize $^1$H spins elsewhere in the system. Increasing the amount of oil has two effects. Firstly, more oil leads to a steady increase in the number of spins in contact with electrons on the ND surface, thus increasing the hyperpolarized signal and $\Delta$S. Secondly, the additional oil in the mixture also increases the signal from spins in thermal equilibrium S$_{\textrm{Th}}$, either at the surface or in the bulk of the liquid. To account for both the hyperpolarized and thermal contributions to the signal, Fig. 5b also shows the enhancement $\epsilon$ = $\Delta$S/S$_{\textrm{Th}}$ as a function of the amount of oil (blue). Similar to the case of the ND-water mixture, we observe a transition in behavior around 60 $\mu$L, where the contribution from hyperpolarization begins to saturate and adding further liquid simply leads to dilution. We speculate that the liquid concentration at which this transition occurs provides information about the packing density of ND, viscosity, and extent of diffusion present in the mixture. Future efforts may exploit such identifiers to signal the adsorption and desorption of ND payloads such as chemotherapeutics. 

\begin{figure}
\includegraphics{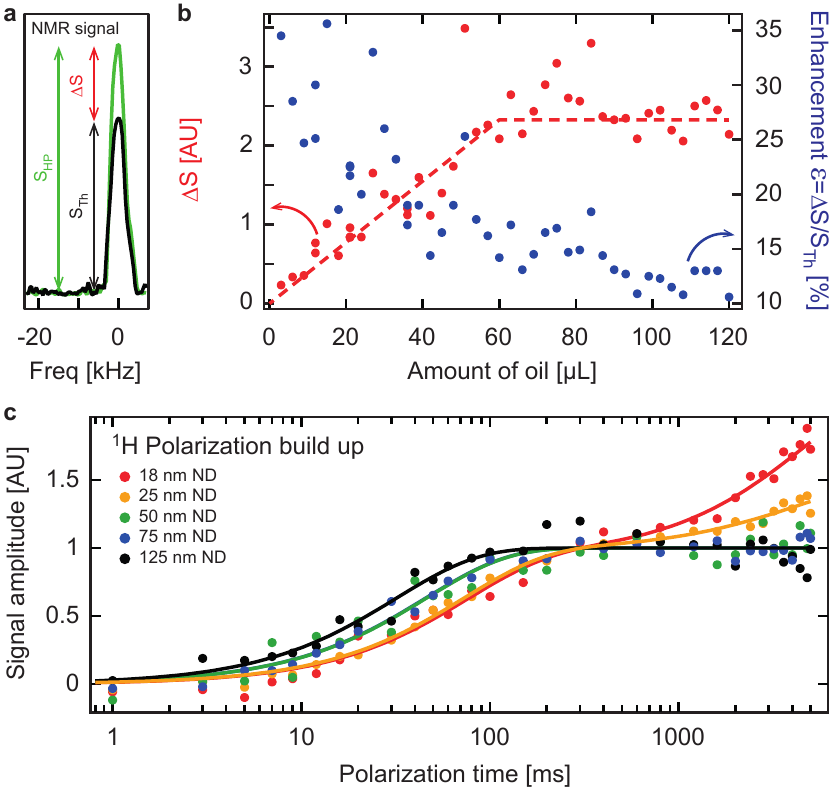}
\caption{\small \textbf{Hyperpolarization dynamics of adsorbed liquids.} {\bf a)} Schematic showing the thermal NMR signal (S$_\textrm{Th}$), the hyperpolarized NMR signal (S$_\textrm{HP}$), and the difference in these signal ($\Delta$S). {\bf b)} Enhancement (blue) and the change in the $^1$H NMR signal with hyperpolarization ($\Delta$S, red) for a ND-oil mixture as a function of oil concentration. Saturation of the ND surface occurs after 60 $\mu$L of oil is added. The data points are the saturation values of polarization build up curves (at $f = \omega_e - \omega_n$). The solid red line is a guide to the eye. {\bf c)} $^1$H Polarization build up at $f = \omega_e - \omega_n$ in an oil-ND mixture for 18 nm ND (red), 25 nm ND (yellow), 50\,nm ND (green), 75 nm ND (blue), and 125 nm ND (black). Solid lines are either exponential fits (50 nm, 75 nm and 125 nm ND) or double exponential fits (18\,nm and 25\,nm ND) to the data. The data has been corrected for heating effects [see methods section]. The polarization build up times are: 18\,nm ND: $\tau_1$\,=\,72\;ms, $\tau_2$\,=\,4.7\;s; 25\,nm\,ND: $\tau_1$\,=\,72\;ms, $\tau_2$\,=\,4.4\;s; 50\,nm ND: $\tau$\,=\,46\;ms; 75\,nm\,ND: $\tau$\,=\,45\;ms; 125\,nm ND: $\tau$\,=\,32\;ms. Data has been normalized such that 0 corresponds to the signal with no microwaves, and 1 corresponds to saturation of the fast component of the polarization build-up.}
\end{figure}

The combined data-sets for water and oil nanodiamond mixtures suggest that hydrogen spins become attached to the ND surface and remain there for times that are long compared with hyperpolarization and relaxation processes. To test this picture further, we examine lastly how the signal enhancement depends on the time over which microwaves are applied to produce hyperpolarization. The signal is observed to grow mostly between 10 and 100\,ms of hyperpolarization, saturating for longer times, as shown in Fig. 5c. This saturation is consistent with the surface adsorbed $^1$H spins undergoing little diffusion into the bulk liquid and thus blocking the surface from being further replenished with new, unpolarized spins. In this regime,  the surface bound spins will reach a steady-state enhancement that is determined by the rates of hyperpolarization and relaxation. Nanodiamonds below 50 nm in size exhibit this saturation in signal, but then undergo a slight further enhancement for hyperpolarization times longer than 1 second. This surprising behavior could be partially explained by diffusion in the oil-ND mixture that becomes enhanced for small diamonds. A further possibility is that the timescale over which $^1$H spins remain adsorbed on the surface is reduced for NDs below a certain size. 

\subsection{Discussion and Conclusion}
The use of nanodiamond in a biological context is now widespread \cite{Say_review}, given they are essentially non-toxic, exhibit a readily functionalized surface as well as attributes that enable new imaging and tracking modalities. Here, we have focused on the spin interactions of the ND surface and liquid interface at room temperature, making strong use of nuclear hyperpolarization to uncover aspects of the dynamics that are otherwise challenging to observe. Beyond a new means of characterization, the use of such hyperpolarization techniques offer a means of detecting the presence or absence of adsorbed compounds, of use in targeted delivery and release of chemotherapeutics. Extending this approach into the spectroscopic domain, either correlating chemical shift or relaxation times, would enable the possibility of distinguishing local environments and compounds interacting with the ND surface. 

Given the long relaxation times  and significant $^{13}$C hyperpolarization that is possible with nanodiamond \cite{us}, the surface spin interactions investigated here open the prospect of transferring polarization from the ND-core to the surface. In this mode, intrinsic surface electrons can act to mediate polarization transfer between $^{13}$C storage and $^1$H spins for detection. Such an approach is amenable to techniques based on microfluidics \cite{NVmicrofluids}. 

In conclusion, we have examined the use of nanodiamond and its surface in establishing polarized states of various liquids at room temperature. The presence of ND leads to enhanced relaxation of $^1$H spins in solution, opening a means of generating ND-specific contrast for MRI. The application of microwaves near the resonance frequency of the surface electrons leads to hyperpolarization of $^1$H spins, consistent with dynamic nuclear polarization via the solid-effect and cross-effect. Finally, the combined use of hyperpolarization and relaxation measurements allow for spins on the ND surface to be distinguished from those in the bulk liquid, opening a means to probe the local environment of ND in vivo.

\subsection{Methods}

{\bf Nanodiamonds.} In these experiments HPHT NDs purchased from Microdiamant were used. We refer to the diamonds by their median size. Measurements were made on MSY 0\nobreak-\nobreak 0.030, (0\nobreak-\nobreak 30\,nm, median 18\,nm), MSY 0\nobreak-\nobreak 0.05 (0-50\,nm, median 25\,nm), MSY 0\nobreak-\nobreak0.1 (0\nobreak-\nobreak100\,nm, median 50\,nm), MSY 0\nobreak-\nobreak 0.15 (0\nobreak-\nobreak 150\,nm, median 75\,nm), MSY 0\nobreak-\nobreak 0.25 (0-250\,nm, median 125\,nm), MSY 0\nobreak-\nobreak 500 (0\nobreak-\nobreak 500\,nm, median 210\,nm), MSY 0.25\nobreak-\nobreak 0.75 (250\,nm\nobreak-\nobreak 500\,nm, median 350\,nm),  MSY 0.25\nobreak-\nobreak 0.75 (250\,nm\nobreak-\nobreak 750\,nm, median 500\,nm), MSY 0.75\nobreak-\nobreak 1.25 (750\,nm\nobreak-\nobreak 1250\,nm, median 1000\,nm) and MSY 1.5\nobreak-\nobreak 2.5 (1500\,nm\nobreak-\nobreak 2500\,nm, median 2000\,nm). 

{\bf Air Oxidation.} HPHT NDs were spread in a thin layer and placed in a furnace at 550$^{\circ}$C for 1 hr (with 1 hr of heating to reach 550$^{\circ}$C and 20 min to return to room temperature). 

{\bf Adsorption.} Initially NDs were heated on a hot plate to remove any adsorbed water. The NDs were mixed with various liquids and sonicated. Adsorption occurred passively. The ND remained suspended in solution for the duration of the experiments.

{\bf Experimental setup.}
Signals were acquired with a single spaced solenoid coil in a home built NMR probe in a magnetic field range of $B$ = 300\;mT - 500\;mT provided by either a permanent magnet ($B$ = 460\;mT) or an electromagnet. X-band microwave irradiation was amplified to a power of 10 W and coupled to the sample using a horn antenna and reflector. NMR signals were measured by initially  polarizing the sample, then detecting the polarized signal using either a $\pi/2$ pulse or an echo ($\pi/2 - \tau- \pi$) sequence, and finally waiting for the polarization to return to thermal equilibrium. Data was acquired using either a Redstone NMR system (Tecmag) or a Spincore NMR system.

{\bf SEM images.} SEM images were taken on a Zeiss Ultra Plus Gemini SEM spectrometer working in transmission mode. 

{\bf Raman spectra.} Raman spectra were acquired with a Renishaw inVia Raman Microscope at $\lambda$ = 488 nm and $P$ = 50 $\mu$W. 

{\bf ESR measurements.} ESR measurements were made using a Bruker EMX-plus X-Band ESR Spectrometer. The cavity Q-factor ranged between $Q$ = 5000 - 10000 for small and large ND particles respectively. ESR power was $P$ = 0.25 $\mu$W, modulation amplitude was 1 Gs, and the modulation frequency was 100 kHz. Simulations of the ESR spectra were performed in Easyspin \cite{easyspin}. Fit parameters were linewidth, signal amplitude and g-factor for each spin species.

{\bf Relaxivity Measurements.} The $T_1$ polarization build up curves were fitted with an exponential fit $M/M_0 = 1-2e^{-t/T_1}$, where $M$ is the magnetization, $M_0$ is the equilibrium magnetization, $T_1$ is the spin lattice relaxation time and $t$ is the polarization build up time. Relaxivity data was fitted with $T_1 = 1/(1/T_{1pure} + R C)$ where $C$ is the concentration of nanodiamond, $T_{1pure}$ is the $T_1$ relaxation time of pure (undoped) water and $R$ is the relaxivity. The water had a T$_1$ relaxation time of T$_{1pure}$ = 2.6 s measured at $B$ = 300 mT.

{\bf Hyperpolarization spectra of ND-liquid mixtures.} For the ND-oil mixtures, approximately 50 mg of ND was mixed with 60 $\mu$L of oil. The mixtures were polarized for 300\,ms, at $B$\,=\,458\,mT and for\,1\,s at other fields in the range $B$\,=\,300\,mT\,-\,500\,mT. Solid lines are double Lorentzian fits to $y = y_0 + \frac{a_1}{(x-x_1)^2 + B_1} + \frac{a_2}{(x-x_2)^2 + B_2}$. 

{\bf T$_1$ relaxation.} ND-oil mixtures: Mixtures of 70 mg of 18 nm ND and de-ionised water were polarized for 300 ms. 

{\bf Enhancement as a function of oil concentration.} 75 mg of 125 nm ND was mixed incrementally with oil. Polarization build up was measured out to 1 second and fitted with an exponential curve. All the curves reached saturation. 

{\bf Polarization build up.} 70 mg of ND was mixed with 40 $\mu$L of oil. With off-resonant microwaves, a signal decrease of 6\% due to heating effects was seen after 1 second of polarization. Data with on-resonant microwaves was corrected to account for this heating effect.

\begin{acknowledgement}

The authors thank T. Boele for useful discussions. For SEM measurements the authors acknowledge the facilities and technical assistance of the Australian Centre for Microscopy \& Microanalysis at the University of Sydney. For Raman measurements the authors acknowledge the staff and facilities at the Vibrational Spectroscopy Core Facility at the University of Sydney. For ESR measurements the authors acknowledge the staff and facilities at the Nuclear Magnetic Resonance Facility at the Mark Wainwright Analytical Centre at the University of New South Wales. This work is supported by the Australian Research Council Centre of Excellence Scheme (Grant No. EQuS CE110001013)  and ARC DP1094439.
\end{acknowledgement}




\providecommand{\latin}[1]{#1}
\providecommand*\mcitethebibliography{\thebibliography}
\csname @ifundefined\endcsname{endmcitethebibliography}
  {\let\endmcitethebibliography\endthebibliography}{}

\end{document}